\documentclass[aps,prd,preprint,tightenlines,groupedaddress,showpacs]{revtex4}
\usepackage{epsf,epsfig,graphics,graphicx}
\begin{document}

\title{Damping of tensor modes in inflation}

\author{Kin-Wang Ng}
\affiliation{Institute of Physics \& Institute of Astronomy and Astrophysics,
Academia Sinica, Taipei 11529, Taiwan}

\date{\today}

\begin{abstract}
We discuss the damping of tensor modes due to anisotropic stress in inflation.
The effect is negligible in standard inflation and may be significantly large
in inflation models that involve drastic production of free-streaming particles.
\end{abstract}

\maketitle

\section{Introduction}

The spatial flatness and homogeneity of the present Universe
strongly suggest that a period of de Sitter expansion or inflation
had occurred in the early Universe~\cite{olive}. During inflation
quantum fluctuations of the inflaton field may give rise to energy
density perturbations (scalar modes)~\cite{pi}, which can serve as
the seeds for the formation of large-scale structures of the Universe.
In addition, a spectrum of gravitational waves (tensor modes) is
produced from the de Sitter vacuum~\cite{star}.

Gravitational waves are very weakly coupled to matter,
so once produced they remain as a stochastic
background till today, and thus provide a
potentially important probe of the inflationary epoch.
Detection of these primordial waves by using terrestrial wave
detectors or the timing of millisecond pulsars~\cite{krau} would indeed
require an experimental sensitivity of several orders of magnitude beyond the current
reach. However, like scalar perturbation,
horizon-sized tensor perturbation induce large-scale temperature
anisotropy of the cosmic microwave background (CMB)
via the Sachs-Wolfe effect~\cite{sach}. The recent
seven-year WMAP anisotropy data has placed an upper limit on the
contribution of tensor modes to the CMB anisotropy, in terms of  the
tensor-to-scalar ratio, that is $r<0.36$~\cite{komatsu}. More stringent
limits, $r<0.17-0.19$, have been made by combining several
other cosmological measurements~\cite{actspt}. In addition, the tensor
modes uniquely induce CMB B-mode polarization that is the primary aim
of on-going and future CMB experiments~\cite{weiss}.

As is well-known, gravitational waves propagate freely in the
expanding Universe~\cite{lifs}. This is under the assumption that the
Universe is a perfect fluid. In the presence of non-vanishing anisotropic
stress, an additional source term to the gravitational wave equation
should be included~\cite{cosmology}.  The effect of anisotropic stress
on cosmological gravitational waves due to free-streaming neutrinos after
the neutrino-matter decoupling in the early Universe
was first numerically calculated in Ref.~\cite{bond}, and incorporated in
an integro-differential equation for the wave propagation~\cite{weinberg}.
In fact, this equation can be also applied for any unknown free-streaming
relativistic particles~\cite{boyle}. It was found that the anisotropic stress
reduces the wave amplitude, thus lowering the tensor-mode induced CMB
anisotropy and polarization~\cite{bond,weinberg,boyle,pritchard}.

In this paper, we will discuss the effect of anisotropic stress on tensor modes
in inflation. Here the anisotropic stress is due to free-streaming relativistic
particles produced during inflation. The generating source of these relativistic particles
could be de Sitter quantum fluctuations of the inflaton itself in standard slow-roll
inflation~\cite{pi}, a thermal component in warm inflation~\cite{berera},
isolated bursts of instantaneous massless particle production~\cite{barnaby},
particle production in trapped inflation in which the inflaton rolls slowly down a
steep potential by dumping its kinetic energy into light particles at the trapping
points along the inflaton trajectory~\cite{green,lee}, or electromagnetic dissipation
in natural inflation~\cite{sorbo,peloso}.

\section{Particle production in inflation}
\label{models}

Here we assume a flat Friedmann-Robertson-Walker metric,
\begin{equation}
ds^2 =dt^2-a^2(t) d{\bf x}^2
     = a^2(\eta) ( d\eta^2-d{\bf x}^2)\,,
\end{equation}
where $a(\eta)$ and $d\eta=dt/a(t)$ are the scale factor and conformal time respectively.
For simplicity, we treat the inflaton energy density $\rho_\phi$ as approximately
constant and then we have
\begin{equation}
a=-\frac{1}{H\eta}\,,\quad
H^2=\frac{8\pi G}{3}\rho_\phi\,.
\end{equation}
We denote the energy density of the free-streaming relativistic
particles $\chi$ produced during inflation by $\rho_\chi$ and define a ratio,
\begin{equation}
f \equiv \frac{\rho_\chi}{\rho_\phi}\,.
\end{equation}

In the standard slow-roll inflation, a candidate for the $\chi$
particle is any weakly interacting field whose quanta
are gravitationally produced during inflation.
These de Sitter quantum fluctuations have a characteristic
energy density roughly equal to $\rho_\chi \sim H^4$~\cite{pi}, so we have
$f\sim GH^2$. The WMAP results~\cite{komatsu} have set an upper limit on the inflation
scale which means that $f<10^{-10}$. We will see below that this small $f$
implies too weak anisotropic stress to affect the tensor modes
in the standard slow-roll inflation.

However, some inflation models that involve particle production via
interactions between $\phi$ and $\chi$ may allow a relatively large value for $f$.
For example, in the trapped inflation model with an interaction of the type
$g^2(\phi-\phi_i)^2\chi_i^2$~\cite{green}, when $\phi$ rolls slowly down a steep potential
by dumping its kinetic energy into $\chi_i$ particles at each trapping point at $\phi_i$
along the inflaton trajectory, a roughly constant energy density of all $\chi_i$'s,
$\rho_\chi=\sum_i \rho_{\chi_i}$, is maintained. This energy density can be estimated as
$\rho_\chi \sim k_*^4$ with $k_*$ being a characteristic energy scale
given by $k_*\sim \sqrt{g (d\phi/dt)}$, where $d\phi/dt$ is the inflaton rolling speed.
To have a successful trapped inflation, it is required that $f<0.1$.
Also, it is shown that the scattering rate of $\chi_i$ particles, which is given by
$\Gamma \sim k_*^4/E^3$ with $E$ being the energy of $\chi_i$,
is sufficiently slower than the expansion rate, $\Gamma \le H$~\cite{senatore}.
Thus, in this model the de Sitter vacuum may be populated with free-streaming
$\chi_i$ particles that generate significantly large anisotropic stress to damp the tensor modes.

Another example that also provides with a constant $\rho_\chi$ during inflation involves
an interaction, $\phi/M F^{\mu\nu}\widetilde{F}_{\mu\nu}$, where $\chi$ is a massless $U(1)$ gauge field,
$F^{\mu\nu}$ is its field strength, and $M$ is a mass scale~\cite{sorbo,peloso}.
The growth solution for the Fourier mode of the vector potential with $+$ circular polarization is found as
\begin{equation}
A_+(\eta,k)\simeq \frac{1}{\sqrt{2k}} \left(\frac{k}{2\xi aH}\right)^{1/4}
e^{\pi\xi -2\sqrt{2\xi k/(aH)}}
\label{Ak}
\end{equation}
in the interval $(8\xi)^{-1} \lesssim k/(aH) \lesssim 2\xi$,
where $\xi \equiv 2(d\phi/dt)/(MH)$ is treated as constant.
Hence, the energy density of the produced gauge quanta is given by
\begin{eqnarray}
\rho_\chi &=& \frac{1}{4\pi^2 a^4}\int dk\,k^2
\left(\left|dA_+/d\eta\right|^2 +k^2\left|A_+\right|^2\right)\nonumber\\
&\simeq& \frac{6!}{2^{19}\pi^2} \frac{H^4}{\xi^3} e^{2\pi\xi}\,.
\label{power}
\end{eqnarray}
These gauge quanta, in turn, source inflaton fluctuations which are
highly nongaussian. The WMAP bound on nongaussianity implies that $\xi\lesssim 3$~\cite{peloso}.
When $\xi= 3$, $f \simeq 6.6\times 10^3 GH^2\ll~1$. However, the value of $\xi$ gets to increase towards
the end of inflation. If $\xi= 5.16$ near the end of inflation and $GH^2=10^{-10}$, then we will have $f\simeq 0.1$.
The gauge quanta, once produced, scatter with the inflaton fluctuations with a rate given by
\begin{equation}
\Gamma \sim \sigma n_{\delta\phi} \lesssim  \frac{H^2}{M^4} \times \frac{\rho_\chi}{H}
\sim 10^{-4}\frac{e^{2\pi\xi}}{\xi^3} \left
(\frac{H}{M}\right)^4 H\,,
\end{equation}
where the energy of the gauge particle is of order $H$ as shown in Eq.~(\ref{Ak}) and
the number density of inflaton fluctuations is approximated as $n_{\delta\phi} \lesssim n_\chi$.
As long as $\xi<5.16$ and $H<10^{-2}M$, we reach the condition $\Gamma<H$, under which
the gauge quanta freeze out and decouple from the background.

\section{Gravitational wave equation}

In the weak field approximation, small metric fluctuations are ripples on the
background metric:
\begin{equation}
g_{\mu\nu}=a^2(\eta)(\eta_{\mu\nu}+h_{\mu\nu})\;,\;\;\;h_{\mu\nu}\ll 1\,,
\label{hmunu}
\end{equation}
where $\eta_{\mu\nu}$ is the Minkowski metric and Greek indices run from 0 to 3.
In synchronous gauge, $h_{00}=h_{0i}=0$, where $i$ runs from 1 to 3. The
remaining $h_{ij}$ contain a transverse, traceless tensor which corresponds to
a gravitational wave or tensor mode. Henceforth we will work in the TT gauge,
i.e., $h^k_k=\partial_i h^{ij}=0$ and denote the two independent polarization states
of the wave as $+$, $\times$.
The propagation of gravitational waves in an expanding space-time is well studied.
In the presence of anisotropic stress, the Fourier mode equation is given by~\cite{cosmology}
\begin{equation}
\frac{d^2 {\tilde h}_{ij}}{d\eta^2}+{2\over a}\frac{da}{d\eta}\frac{d{\tilde h}_{ij}}{d\eta}+k^2 {\tilde h}_{ij}
=16\pi G a^2 \pi_{ij}\,,
\label{modeeq}
\end{equation}
where $\pi_{ij}$ is the Fourier mode of the TT part of the anisotropic stress tensor.

\section{Anisotropic stress tensor}

In this section, we will derive the evolution of the anisotropic stress tensor of the
free-streaming relativistic particles $\chi$. We will follow the methodology
in Ref.~\cite{cosmology}, taking into account the particle production in inflation
and assuming that the produced $\chi$ particles are decoupled from the background.
Let $N_\chi ({\bf x}, {\bf p}, t)$ be the phase space density of $\chi$ particles,
then the physical energy density of $\chi$ particles is given by
\begin{equation}
\rho_{\chi}(t)= a^{-4}(t){1\over V}\int_V d^3{\bf x}\, d^3{\bf p}\, |{\bf p}| N_\chi ({\bf x}, {\bf p}, t).
\label{Echi}
\end{equation}
In light of the results in Sec.~\ref{models}, in the following we will assume that the physical energy density
is constant during inflation: $d\rho_{\chi}(t)/dt =0$. This suggests that we should deal with a re-scaled phase space density instead, defined by
\begin{equation}
n_\chi ({\bf x}, {\bf p}, t)\equiv a^{-4}(t) N_\chi ({\bf x}, {\bf p}, t).
\end{equation}

In the absence of collisions, the re-scaled phase space density
satisfies a Boltzmann equation in a metric $g_{ij}({\bf x}, t)$,
\begin{equation}
\frac{\partial n_\chi}{\partial t} + \frac{\partial n_\chi}{\partial x^i} \frac{p^i}{p^0}
+\frac{\partial n_\chi}{\partial p_i} \frac{p^j p^k}{2p^0} \frac{\partial g_{jk}}{\partial x^i}=0,
\label{bzeq}
\end{equation}
where $p^i=g^{ij}p_j$ and $p^0=\sqrt{g^{ij}p_i p_j}$. We now consider small perturbation,
\begin{equation}
g_{ij}=a^2\delta_{ij}+\delta g_{ij},\quad n_\chi={\bar n}_\chi (a p^0)+\delta n_\chi,
\end{equation}
where ${\bar n}_\chi (p)$ with $p=|{\bf p}|$ is just the re-scaled local phase space density,
which has a well-defined energy spectrum as exemplified in Eq.~(\ref{power}).
To first order in metric and density perturbation,
Eq.~(\ref{bzeq}) reads
\begin{eqnarray}
&&\frac{\partial{\bar n}_\chi (p)}{\partial t}=0,\\
&&\frac{\partial \delta n_\chi}{\partial t} + \frac{{\hat p}_i}{a} \frac{\partial \delta n_\chi}{\partial x^i}
={p\over 2}\frac{\partial {\bar n}_\chi (p)}{\partial p}{\hat p}_i {\hat p}_j  \frac{\partial}{\partial t}
   \left(\frac{\delta g_{ij}}{a^2}\right), \label{pertbzeq}
\end{eqnarray}
where ${\hat p}_i=p_i/p$.

Using Eq.~(\ref{hmunu}) for the metric perturbation, we write down the tensor component of Eq.~(\ref{pertbzeq}) as
\begin{equation}
\frac{\partial \delta n_\chi}{\partial t} + \frac{{\hat p}_i}{a} \frac{\partial \delta n_\chi}{\partial x^i}
={p\over 2}\frac{\partial {\bar n}_\chi (p)}{\partial p} {\hat p}_i {\hat p}_j \frac{\partial h_{ij}}{\partial t}.
\label{boltzn}
\end{equation}
Let us introduce a dimensionless re-scaled intensity perturbation defined by
\begin{eqnarray}
{\bar\rho}_{\chi} J ({\bf x}, {\hat p}, t) &\equiv& \int dp\, 4\pi p^3\, \delta n_\chi ({\bf x}, {\bf p}, t), \\
{\bar\rho}_{\chi} &\equiv& \int dp\, 4\pi p^3 {\bar n}_\chi (p).
\end{eqnarray}
Then, Eq.~(\ref{boltzn}) becomes
\begin{equation}
\frac{\partial J}{\partial t} + \frac{{\hat p}_i}{a} \frac{\partial J}{\partial x^i}
=-2{\hat p}_i {\hat p}_j \frac{\partial h_{ij}}{\partial t},
\end{equation}
where we have used integration by part and assumed that ${\bar n}_\chi (0)={\bar n}_\chi (\infty)=0$.
We can then construct the spatial component of the stress tensor perturbation as
\begin{equation}
\delta T_{\chi j}^i = \int d^3 {\bf p}\, \delta n_\chi ({\bf x}, {\bf p}, t)\,p {\hat p}_i {\hat p}_j,
\end{equation}
which contributes to the anisotropic stress tensor $\pi_{ij}$.
Following the same steps in Ref.~\cite{cosmology}, the free-streaming solution for the anisotropic 
stress tensor of $\chi$ particles in the presence of gravitational waves is found as
\begin{equation}
\pi^{\chi}_{ij}= -4{\bar\rho}_{\chi}\int_{\eta_i}^{\eta} K(k\eta-k\eta')\frac{d {\tilde h}_{ij}(\eta')}{d\eta'}\, d\eta'\,,
\label{piij}
\end{equation}
where $\eta_i$ is some initial time and the kernel is given by
\begin{equation}
K(u)=\frac{j_2(u)}{u^2}=-\frac{\sin u}{u^3}-\frac{3\cos u}{u^4}+\frac{3\sin u}{u^5}\,.
\end{equation}
The integro-differential equation~(\ref{modeeq}) with $\pi_{ij}$ given by Eq.~(\ref{piij})
has been solved for the case in which $\chi$ particles are free-streaming neutrinos in the early Universe~\cite{weinberg,boyle,pritchard}.
One would anticipate that the free-streaming solution of the anisotropic stress
tensor~(\ref{piij}) is a back reaction to the wave equation and thus reducing the
wave amplitude.

In trapped inflation~\cite{green} or axionic inflation with $10^5$ $U(1)$ gauge
fields~\cite{sorbo}, the time-scale of particle production is much shorter than
the expansion time, $H^{-1}$. For instance, trapped inflation produces particles
in a time-scale of order $k_*^{-1} \ll H^{-1}$. Therefore, inflation begins very shortly
after particles are copiously produced. Let $\eta_i$ be the moment when inflation begins.
The initial condition, $\pi^{\chi}_{ij}(\eta_i)=0$, that we have assumed in Eq.~(\ref{piij})
is then justified.

Note that generically anisotropic stress should exist before inflation, i.e. $\pi_{ij}(\eta_i)\neq 0$,
due to the fact that we do not really know the initial condition for inflation since here we do not have a physical
model before inflation. However, soon after inflation starts, this pre-existing
anisotropic stress has decayed and become vanishingly small. Since we are mainly
interested in $\chi$ particles produced during inflation, we have assumed that the
generic anisotropic stress is absent at the beginning of inflation, namely $\pi_{ij}(\eta_i)=0$.
Otherwise, we will need to consider the effect of this generic anisotropic stress on
gravitational waves in a brief period after the start of inflation.

\section{Damping in inflation}

Let us decompose
\begin{equation}
{\tilde h}_{ij}(\eta, {\bf k})=(8\pi G)^{\frac{1}{2}} H k^{-\frac{3}{2}} h(\eta,{\bf k})
\epsilon_{ij}({\bf k};\lambda)\,,
\end{equation}
where $\epsilon_{ij}({\bf k};\lambda)$ is the polarization tensor with $\lambda=+,\times$.
Here we have introduced a dimensionless wave amplitude $h$ that is assumed
to be the same for both polarizations. Then, Eq.~(\ref{modeeq}) becomes
\begin{equation}
\frac{d^2 h}{du^2}-\frac{2}{u}\frac{d h}{du}+h
=-\frac{24 f}{u^2}\int_{u_i}^{u} K(u-u') \frac{dh}{du'}\, du'\,.
\label{hueq}
\end{equation}
where $f={\bar\rho}_{\chi}/\rho_\phi$ is a constant.
The homogeneous solution of Eq.~(\ref{hueq}) is known. Selecting the
Bunch-Davis vacuum~\cite{BD}, it is given by
\begin{equation}
h_0=-(1+iu)\, e^{-iu}.
\end{equation}
At the end of inflation (i.e., $u\rightarrow 0$), $|h_0|\rightarrow 1$. This reproduces
the scale-invariant power spectrum predicted in standard slow-roll inflation.

\begin{figure}[htp]
\centering
\includegraphics[width=0.8\textwidth]{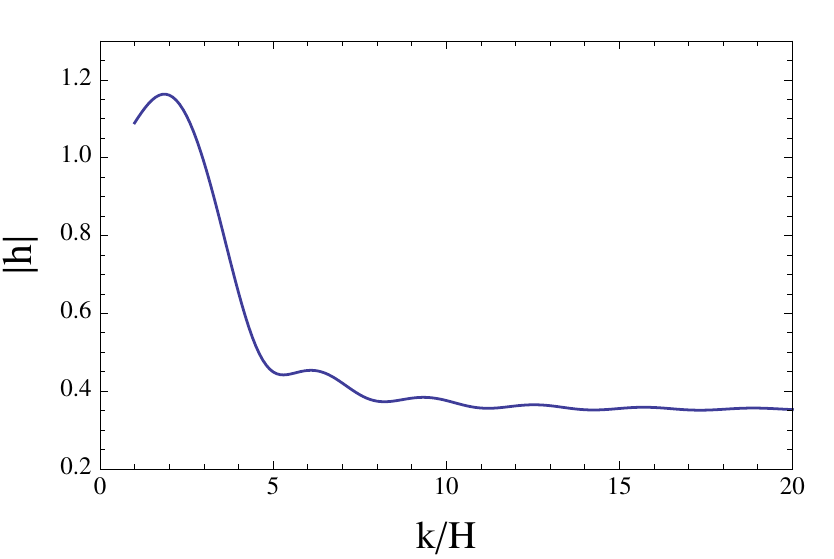}
\caption{The wave amplitude $|h|$ versus the wave number $k/H$ at the end of inflation
by the use of the Born approximation in Eq.~(\ref{hp}),
where $24f=1$ is assumed and $k/H=1$ corresponds to the length scale that leaves the horizon
at the onset of inflation.}
\label{fig1}
\end{figure}

Since the damping effect is expected to be secondary, we can make the Born
approximation to replace $dh/du'$ in the damping term in Eq.~(\ref{hueq})
by $dh_0/du'$. Then, we use the retarded
Green's function method to find the particular solution,
\begin{equation}
h_p=-24f\int_{u_i}^u \frac{du'}{u'^2} G(u-u')
\int_{u_i}^{u'} du'' K(u'-u'') \frac{dh_0}{du''}\,,
\label{hp}
\end{equation}
where the Green's function is constructed from the homogeneous solution, given by
\begin{equation}
G(u-u')=\frac{u}{u'}\left[\left(1+\frac{1}{uu'}\right)\sin(u-u')-
\frac{u-u'}{uu'}\cos(u-u')\right].
\end{equation}
Now we set $a=1$ at the time when inflation starts. This fixes $u_i= -k/H$ with
$k/H=1$ corresponding to the length scale that crosses the horizon at the onset of
inflation. We then numerically evaluate the integral~(\ref{hp}) by letting
$u\rightarrow 0$ to obtain $h_p$ for a range of values of $k/H$. In Fig.~\ref{fig1},
assuming that $24f=1$, we plot the total wave amplitude, $|h|=|h_0+h_p|$,
at the end of inflation ($u\rightarrow 0$) against the wave number, $k/H$.
The figure shows that the wave amplitude is reduced and becomes asymptotically flat at large $k$.

Apparently, the wave amplitude is enhanced for $k/H<3$. However, we do not expect that the
present work would give precise results for these low $k$-modes. The enhancement should be an artifact
due to the use of the approximation, $dh/du'\simeq dh_0/du'$, and the choice of the Bunch-Davis
vacuum near the initial time, $u_i$. A quick way to resolve this problem, albeit artificial, is to integrate
the damping effect on each $k$-mode for the time interval from the start of inflation to the horizon crossing
time, which means by setting $u= -1$ in Eq.~(\ref{hp}). As such, $|h|=1$ for $k/H=1$ by default.
Some improvements can also be made such as using an advanced solving method for the integro-differential
equation~(\ref{modeeq}) and considering a smooth transit to the inflationary phase.

\section{Conclusion}

We have discussed the damping effect of anisotropic stress on tensor modes
due to free-streaming relativistic particles produced during inflation.
The damping increases with the ratio of the particle energy density and
the de Sitter vacuum energy, which ranges from $10^{-10}$ in standard inflation
to about $0.1$ in inflation models that involve drastic particle production
such as the trapped inflation or axionic inflation. In these inflation models,
the particle production may significantly reduce the amplitude of the tensor modes.

Recently new sources of gravitational waves during inflation have been proposed.
They are anisotropic stress induced by quantum energy stress of conformal fields~\cite{hsiang}
and their associated fluctuations~\cite{wu},
by the Bremsstrahlung from the particle production events~\cite{senatore},
and by the produced gauge field quanta that couple to inflaton~\cite{sorbo11,cook}.
All of these produced gravitational waves may experience the damping effect considered
in the present work if copious free-streaming relativistic particles are also produced
during inflation. As such, to properly take into account the damping, one needs to consider the
full integro-differential equation,
\begin{equation}
\frac{d^2 {\tilde h}_{ij}}{d\eta^2}+{2\over a}\frac{da}{d\eta}\frac{d{\tilde h}_{ij}}{d\eta}+k^2 {\tilde h}_{ij}
+64\pi G a^2{\bar\rho}_{\chi}\int_{\eta_i}^{\eta} K(k\eta-k\eta')\frac{d {\tilde h}_{ij}(\eta')}{d\eta'}\, d\eta'
=16\pi G a^2 \pi_{ij}^{\rm new}\,,
\label{fulleqn}
\end{equation}
where the damping term is taken from  Eq.~(\ref{piij}) and $\pi_{ij}^{\rm new}$ denotes a new source term for the anisotropic stress.
When $f\ll 1$, the damping term can be neglected and the equation reduces to that
considered in Refs.~\cite{hsiang,wu,senatore,sorbo11,cook}.
Otherwise, Eq.~(\ref{fulleqn}) should be
solved self-consistently to obtain the damped tensor power spectrum.
At last we note that in Ref.~\cite{cook} they have considered
the production of gravitational waves at the interferometric scales during the final
stage of inflation when $\xi\sim 5-6$. This large value of $\xi$ may imply that $f\sim 1$,
thus resulting in a large damping on the tensor modes.

\begin{acknowledgments}

The author would like to thank E. Pajer, L. Senatore, and E. Silverstein for useful
discussions, and Stanford Institute for Theoretical Physics for the hospitality,
where a part of the work was done.
This work was supported in part by the National
Science Council, Taiwan, ROC under Grant No.
NSC98-2112-M-001-009-MY3.

\end{acknowledgments}

\end{document}